\documentclass[a4paper,11pt]{article}
\usepackage[british]{babel}
\usepackage[top=1.5cm,bottom=1.5cm,left=1.5cm,right=1.5cm]{geometry}
\usepackage{graphicx}
\usepackage[hyphens]{url}
\usepackage{paralist}
\usepackage{authblk}
\usepackage{siunitx}
\usepackage[noadjust]{cite}
\usepackage[pdftex,colorlinks=true]{hyperref}

\title{Reproducibility in Research: Systems, Infrastructure, Culture}
\author[1]{Tom Crick}
\author[2]{Benjamin A. Hall}
\author[3]{Samin Ishtiaq}
\affil[1]{Cardiff Metropolitan University, UK}
\affil[2]{University of Cambridge, UK}
\affil[3]{Microsoft Research Cambridge, UK}
\affil[1]{\protect\url{tcrick@cardiffmet.ac.uk}}
\affil[2]{\protect\url{bh418@mrc-cu.cam.ac.uk}}
\affil[3]{\protect\url{samin.ishtiaq@microsoft.com}}

\date{ }

\begin{document}
\maketitle

% Keywords:
% reproducible research; e-infrastructure; scientific workflows; computational
% science; open science; data sharing; code sharing; best practices

\begin{abstract}
The reproduction and replication of research results has become a
major issue for a number of scientific disciplines. In computer
science and related computational disciplines such as systems biology,
the challenges closely revolve around the ability to implement (and
exploit) novel algorithms and models. Taking a new approach from the
literature and applying it to a new codebase frequently requires local
knowledge missing from the published manuscripts and transient project
websites. Alongside this issue, benchmarking, and the lack of open,
transparent and fair benchmark sets present another barrier to the
verification and validation of claimed results.

In this paper, we outline several recommendations to address these
issues, driven by specific examples from a range of scientific
domains.  Based on these recommendations, we propose a high-level
prototype open automated platform for scientific software development
which effectively abstracts specific dependencies from the individual
researcher and their workstation, allowing easy sharing and
reproduction of results. This new e-infrastructure for reproducible
computational science offers the potential to incentivise a culture
change and drive the adoption of new techniques to improve the quality
and efficiency -- and thus reproducibility -- of scientific
exploration.
\end{abstract}

\section{Introduction}

Marc Andreessen (co-author of Mosaic, the first widely used web
browser) boldly stated in 2011 that ``{\emph{software is eating the
world}}''~\cite{andreessen:2011}. This is true: we live in a
computational world, with our everyday communications, entertainment,
shopping, banking, transportation, national security, etc, all heavily
data-driven and largely overtaken by software.

Andreessen's statement is particularly true for science and
engineering. A 2012 report by the UK's Royal Society stated that
computational techniques have ``{\emph{moved on from assisting
scientists in doing science, to transforming both how science is done
and what science is done}}''~\cite{rssaaoe:2012}. Many of the examples
discussed in this paper exploit a fundamental advantage of computer
science and more generally, computational science: the unique ability
for researchers to share the raw outputs of their work as software and
datafiles. New experiments, simulations, models, benchmarks, even
proofs increasingly cannot be done without software. This software
does not consist of simple hack-together, use-once, throw-away
scripts; research software repositories contain thousands, perhaps
millions, of lines of code and they increasingly need to be actively
supported and maintained. More importantly, with reproducibility being
a fundamental tenet of science, they need to be open and re-useable.

However, if we closely analyse the scientific literature related to
software tools it often does not appear to be adhering to these
rules~\cite{nature:2011,alberts-et-al:2015}. How many of them are open
and available? How many explain their experimental methodologies, in
particular the basis for their benchmarking? In particular, can we
(re)build the code?~\cite{collberg+proebstring:2016} We, the authors, are
perhaps as guilty as anyone in the past, where we have published
papers~\cite{crick-et-al:2009a,Berdine2011SLAyer} with benchmarks and
promises of code to be released in the near future which depreciate as
you move onto the next project.

There are various reasons why the wider scientific community is in
this position. We are currently undergoing significant changes to
models of academic dissemination, especially considering the wider
open research movement, with new models being
proposed~\cite{deroure:2010,stodden-et-al:2013,fursin+dubach:2014}. Now,
numerous ``high-impact'' journals explicitly require that source code
and data is made available online under some form of open source
license, but there still exists large disciplinary gaps. While these
initiatives are great, they are often optional, seem piecemeal, and do
little to enable the verification and validation of scientific results
at a later stage. Even within the same field, there are different
ideas of what defines reproducibility~\cite{nasem:2016}, as well as
evidence of ``overturn bias'' -- replications that overturn original
results are much easier to publish than those that confirm original
results~\cite{galiani-et-al:2017}.

Nevertheless, the reproduction and replication of reported scientific
results has become a widely discussed topic within the scientific
community~\cite{barnes:2010,morin-et-al:2012,joppa-et-al:2013}.
Whilst the increasing number of retractions of studies across a
variety of disciplines has drawn the focus of many commentators,
automated systems, which allow easy reproduction of results, offer the
potential to improve the efficiency of scientific exploration and
drive the adoption of new techniques. However, just publishing
(linked) scientific data is not enough to ensure the required
reusability~\cite{bechhofer-et-al:2013}. There exists a wider
socio-cultural problem that pervades the scientific community, with
estimates that as much as 50\% of published studies, even those in
top-tier academic journals, cannot be repeated with the same
conclusions by an industrial
lab~\cite{osherovich:2011,hesman-saey:2015}. There are numerous
non-technical impediments to making software maintainable and
re-useable. The pressure to ``make the discovery'', publish quickly
and move onto the next project disincentivises careful software
curation and preservation. Releasing code prematurely is often seen to
give your competitors an advantage, but we should be shining light
into these ``black boxes''~\cite{morin-et-al:2012}; in essence: better
software, better research~\cite{goble:2014}.

However, there is promising existing work in this
area~\cite{chirigati-et-al:2013,stodden+miguez:2014,stodden-et-al:2015,stodden-et-al:2016},
with a variety of manifestos for reproducible research and community
initiatives~\cite{rrspecissue:2008,rrcise:2010,gent:2013,fursin-et-al:2014,Bailey_setthe,james-et-al:2014},
top tips and ``ten simple
rules''~\cite{prlic+proctor:2012,masum-et-al:2013,sandve-et-al:2013,osborne-et-al:2013,goodman-et-al:2014,crick-et-al-irreprod:2015,list-et-al:2017},
as well as analysis of the wider legal, professional, ethical and risk
perspectives~\cite{stodden:2008,haas:2016}. Things can, should and
need to be much better if we want to uphold and maintain the
scientific tenets of openness and sharing. Building upon previous
work~\cite{crick-et-al_wssspe2,crick-et-al_recomp2014}, we present a
call to action, along with a set of recommendations which we hope will
lead to better, more sustainable, more re-useable software, to move
towards an imagined future practice and usage of scientific software
development. We also propose a high-level specification for a service
that would automate many of our recommendations.
% along with curated recommendations on where to publish research
% software (\url{http://www.software.ac.uk/resources/guides/which-journals-should-i-publish-my-software})

\section{We Need to Talk About Reproducibility}

% Recommendation I
\subsection{{\emph{Can I Implement Your Algorithm?}}}

Reproducibility is a fundamental tenet of high-quality research. Yet
many descriptions of algorithms are too high-level, too obscure, too
poorly-defined to allow an easy re-implementation by a third party. A
step in the algorithm might say: ``{\emph{We pick an element from the
frontier set}}'' but which element do you pick? Will the first one do?
Why will any element suffice? Sometimes the author would like to give
more implementation detail but is constrained by an arbitrary page
limit of a conference or journal paper. Sometimes the authors'
description in-lines other algorithms or data structures that perhaps
only that author is familiar with.

Until recently, reproducibility was only discussed at conferences and
workshops convened explicitly for that purpose. This is changing, and
a number of high-profile computer science venues such as the ACM
SIGPLAN conferences POPL and PLDI now explicitly acknowledge the
importance of reproducibility, promoting community-driven reviewing
and validation of software artefacts. \\

\noindent {\textbf{Recommendation {\textrm{I}}:}} We recommend that a
paper must describe the algorithm in such a way that it is
implementable by any knowledgeable reader of that algorithm. The
description is, of course, subjective, but to help encourage better
descriptions, we also recommend that --- in addition to having
incentives to support sharing of computational artefacts --- relevant
scientific conferences develop special tracks for papers that
re-implement past papers' algorithms, techniques or tools.

% Recommendation II
\subsection{{\emph{Set The Code Free}}}

There can be no better proof of your algorithm working, than if you
provide the source code of an implementation; software development is
hard, but sharing and re-using code is relatively easy.

Many years ago, Richard Stallman (founder of the GNU Project and Free
Software Foundation) postulated that all code would be
free~\cite{rms:2010} and we would make our money by consulting on the
code.  As it turns out, this is now the case for a significant
part of the computing industry. There are, of course, hard commercial
pressures for keeping code closed-source. Even in the scientific
domain, scientists and their collaborators may wish to hold onto their
code as a competitive advantage, especially if there exists larger
competitors who could use the available code to ``reverse scoop'' the
inventors, charging into a promising new research area opened by the
inventors.

Closed source is one thing; licenses that deny the user from viewing,
modifying, or sharing the source are another thing. There are,
however, even licences on widely adopted tools like
Gaussian~\cite{Giles2004} (for computational chemistry) that prohibit
even analysing software performance and behaviour. For example, a wide
variety of licenses exist for molecular dynamics software, with
different degrees of openness e.g. Gromacs uses the GNU Lesser General
Public License (LGPL)~\cite{Hess2008}, CHARMM and Desmond are
Academic/Commercial software licences~\cite{Brooks2009,Bowers2006},
Amber and NAMD are custom open-like licences. Z3 is an example from
the verification area: the code itself was only recently open
sourced, but the previous MSR-LA license allowed the source code to be
read, copied, forked for academic use, providing researchers in the
field substantial flexibility~\cite{deMoura2012Z3open}.

Even ignoring licensing issues, sometimes the source is not made open
because the author thinks that it is not quite finished.  You should
follow the ``release early, release often'' mantra, as well as
releasing somewhere public like GitHub, where it is easy to share and
fork. Your code is good enough~\cite{barnes:2010}.
\\
 
\noindent {\textbf{Recommendation {\textrm{II}}:}} There is little
doubt that, if scientific research wants to be open and free, then the
code that underlies it too needs to be open and free. Code that is
available for browsing, modifying, and forking, facilitates testing
and comparison.  We recommend that code be published under an
appropriate open source license~\cite{osl}; while we defer legal
discussion of the specifics of any particular licences, BSD and Apache
are good, flexible ones.\\
% IANAL...

% Recommendation III
\subsection{{\emph{Be A Better Academic Citizen}}}

If you have the appropriate knowledge, skills and experience, you can
create better software. We have seen the emergence of successful
initiatives, such as the Software Sustainability Institute
(\url{http://www.software.ac.uk}) and the UK Community of Research
Software Engineer (\url{http://www.rse.ac.uk}), in cultivating
world-class research through software, developing software skills and
raising the profile of research software engineers.

Many scientists will not have had any formal, or even informal,
training in scientific software development. Building upon the work of
Software Carpentry (\url{http://software-carpentry.org}) and Data
Carpentry (\url{http://www.datacarpentry.org}), basic training in
software engineering concepts like version control (git, mercurial),
unit testing (tests written to exercise the smallest testable parts of
a system, like a function exported from a module), regression testing
(a test framework that ensures that previous results are maintained
over the changes in the source code), build tools (Make, scons), etc,
can help improve the quality of the software written
enormously~\cite{wilson2006}.  Interestingly, many of these concepts
are taught to computer science undergraduates, but it could be argued
that they are taught at the wrong time of their careers, without the
experience of complex, long-running projects.\\
%\footnote{\url{http://philipwfowler.wordpress.com/2013/12/19/the-oxford-software-carpentry-boot-camp-one-year-on/}}.

\noindent {\textbf{Recommendation {\textrm{III}}:}} Software
development skills should be regarded as fundamental literacies for
scientists and engineers: we recommend that formal programming, data
and computational skills are taught as core at undergraduate and
postgraduate level.

% Recommendation IV
\subsection{{\emph{The Lingua Franca of Computational Research}}}

There is no other scientific or technical field where its participants
can just make up a non-principled artefact like a programming language
so easily. In a way, it shows how much of a ``commons'' computer
science has become, that anyone can create a new programming language,
API, framework or compiler. This clearly has its advantages and
disadvantages.

High-level languages are generally more readable than their low-level
relations. The ``density'' of a program is often seen to be a good
thing, but it is not always the case that a shorter Haskell program
(for example) is easier to maintain than a longer C++ one; what is
important is the readability of the code itself. A good example here
is from the world of automatic theorem proving: the SSReflect language
is much more readable than the original, standard Coq
language~\cite{GonthierZND13}. SSReflect uses mathematicians'
vernacular for script commands, allows reproducibility of automatic
proof-checking because parameters are named rather than numbered.
Even though these proof scripts are really only ever going to be run
by a machine, they seek to maintain the basic mathematical idea that a
proof should be readable by another mathematician.

Many high-level programming languages impose constraints like types: that
you can never add a number and a string is the most basic example, but
ML's functors provide principled ways of plugging in components with
their implementations completely hidden. Aggressive type checking
avoids a subset of bugs which can arise due to incorrectly written
functions e.g. well publicised problems with a NASA Mars
orbiter (\url{http://www.cnn.com/TECH/space/9909/30/mars.metric.02/}).
A further example is a pressure coupling
bug (\url{http://redmine.gromacs.org/issues/14}) in
Gromacs~\cite{Hess2008}, which arose due to the inappropriate swapping
of a pressure term with a stress tensor.  A further extension of
types, a concept called units of measure that is implemented in
languages such as F\#, can deal with these kinds of bugs at compile
time. Similarly, problems found using in-house software for
crystallography led to the retraction of five
papers~\cite{Miller2006}, due to a bug which inverted the phases.\\

\noindent {\textbf{Recommendation {\textrm{IV}}:}} The use of a
principled, high-level, typed programming language in which to write
your software helps hugely with the maintainability, robustness and
openness of the software produced. Even in web frontend work, you have
choices: use Typescript or Flow rather than plain old Javascript; use
Hack rather than PHP.

% Recommendation V
\subsection{{\emph{Lineage (or: ``Standing On The Shoulders Of Giants'')}}}

Research software is not just software -- it is the instantiation of
novel algorithms and data structures (or at least novel applications
of data structures). Thus, lineage is important:\\

\noindent {\textbf{Recommendation {\textrm{V}}:}} Code should always
include links to papers publishing key algorithms and the code should
include explicit relationships to other projects on the repository
(i.e. {\emph{Project B}} was branched from {\emph{Project A}}). This
ensures that both the researchers and software developers working
upstream of the current project are properly credited, encouraging
future sharing and development. Remember, the people who did the
research are not necessarily the same people as the developers and
maintainers of the software, so it is important to reward both
appropriately with citations: take note of the FORCE11 Software
Citation Principles~\cite{smith-et-al:2016}.

% Recommendation VI
\subsection{{\emph{YMMV}}}

\begin{figure}[!ht]
\centering
\includegraphics[width=0.8\textwidth]{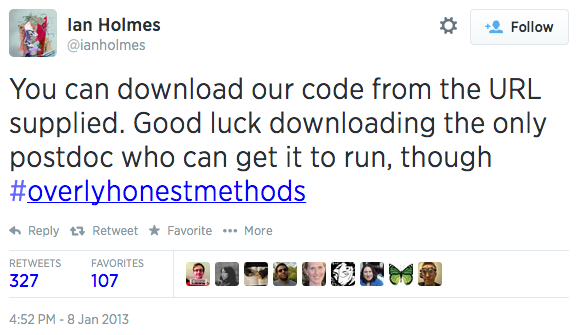}
\caption{{\texttt{\#overlyhonestmethods}} on Twitter by @ianholmes\newline [source: \url{https://twitter.com/ianholmes/status/288689712636493824}]}
\label{fig:overlyhonestmethod} 
\end{figure}

% (\url{http://www.phdcomics.com/comics.php?f=1689})
The tweet in Figure~\ref{fig:overlyhonestmethod} is satirical but
worryingly true, highlighting the perils of reproducible
research. Often, the tool that the paper describes does not exist for
download. Or runs only on one particular bespoke platform. Or might
run for the author, for a while, but will `bit-rot' so quickly that
even the author cannot compile it the following year. Computational
reproducibility would appear to be more straightforward than
replicating physical experiments, but the complex and rapidly changing
nature of computer systems and environments that are being used across
different disciplines makes being able to reproduce and extend such
work a serious challenge~\cite{boettiger:2015}.\\

\noindent {\textbf{Recommendation {\textrm{VI}}:}} You must provide
the source code of the tool, but also with details of precisely
\emph{how} you built and wrote the software. For example:

\begin{itemize}
\item you should provide the compiler and build toolchain; 
\item you should provide build tools (e.g. Makefiles/Ant/etc) and
  comprehensive build instructions; 
\item you should list or link to all non-standard packages and libraries that you use; 
\item you should note the specifics of the hardware and OS used. 
\end{itemize}

This may appear to be significant extra overhead for researchers, but
GitHub APIs, continuous integration servers, virtual machines and
cloud environments can make it easier; see
Section~\ref{sec:conclusion} for more on this.

% Recommendation VII
\subsection{{\emph{Data Representations and Formats}}}

We often do not, and should not, care how things are stored on disk,
what their precise representations are. A common, constrained,
standard representation is however good for passing tests or models around
between different tools. A properly described representation, like the
SMT-LIB format (\url{http://smt-lib.org}) for Satisfiability
Modulo Theory (SMT) solvers, where both the syntax and semantics are
well understood, hugely aids developing tools, techniques and
benchmarks.

Another example, from biology, is that of the standard representation
of qualitative networks and Boolean
networks~\cite{Kauffman1969,Schaub2007}.  These networks can be
expressed in SMV format, but this would mean that standard
qualitative/Boolean network behaviours have to be hard-coded for each
variable, introducing the possibility for errors. In the
BioModelAnalyzer tool~\cite{Benque2012}, the JSON contains \emph{only}
the modifiable parameters limiting the possibility for error; the 
SBML-Qual standard achieves a similar goal for logical models \cite{Chaouiya2013}.\\

% cite XKCD? https://xkcd.com/927/
\noindent {\textbf{Recommendation {\textrm{VII}}:}} Avoid creating
new representations when common formats already exist. Use existing
extensible internationally standardised representations and formats to
facilitate sharing and re-use.

% Recommendation VIII
\subsection{{\emph{World Records}}}

The benchmarks the tool describes are fashioned only for this instance
of this time. They might claim to be from the Microsoft Windows device
driver set, but the reality is that they are stripped down versions of
the originals. Stripped down so much as to be useless to anyone but
the author vs. the referee. It is worse than that really: enough
benchmarks are included to beat other tools. The comparisons are never
fair (especially for comparisons against your tool). If every paper
has to be novel, then every benchmark, too, will be novel; there is no
monotonic, historical truth in new, synthetically-crafted
benchmarks. It is as if, in order to beat Usain Bolt's
\num{100}\si{\metre} world record time, you make him race overweight
and out of season, with a winter overcoat and the wrong sized
shoes. Given this setup, you could surely hope to beat his
\num{9.58}\si{\second} time on a shorter length track.\\

\noindent {\textbf{Recommendation {\textrm{VIII}}:}} Benchmarks should
be public. They should allow anyone to contribute, implying that the
tests are in a standard format. Further, these benchmarks must be
heavily curated. Every test/assertion should be justified. Papers
should be penalised if they do not use these public benchmarks. While
there are some domains in which it may not be immediately possible to
share full benchmarks sets, this should be the exception (with
justification) rather than the norm.\\

A good example of some of these points is the RCSB Protein Data
Bank (\url{http://www.pdb.org}) and Systems Biology Markup
Language~\cite{Chaouiya2013}. The software ones we know of,
the SMT
Competition (\url{http://smtcomp.sourceforge.net/2014/}),
SV-COMP (\url{http://sv-comp.sosy-lab.org/2015/}) and
Termination Problems Data
Base (\url{http://termination-portal.org/wiki/TPDB}) are on
that journey. Such repositories would allow the tests to be taken and
easily analysed by any competitor tool. Some communities go further;
the Critical assessment of methods of protein structure prediction
and prediction of interactions (CASP and CAPRI)\cite{Moult2015,Lensink2017} communities present 
a single-blind test of protein folding and docking
algorithms annually, allowing open competition on a level playing 
field. Similarly the DREAM challenges (\url{http://dreamchallenges.org/}) 
attempt to address large scale problems through open competition.

% Recommendation IX
\subsection{{\emph{Test It To See}}}

Some models may be chaotic and influenced by floating-point errors
(e.g. molecular dynamics), further frustrating testing. For example:
Sidekick is an automated tool for building molecular models and
performing simulations~\cite{Hall2014Sidekick}. Each system is
simulated from an different initial random seed, and under most
circumstances this is the only difference expected between
replicas. However, on a mixed cluster with both AMD and Intel
microprocessors on the nodes, the difference in architecture was found
to alter the number of water molecules added to each system by
one. This meant that the same simulation performed on different
architectures would diverge. Similarly, in a different simulation
engine, different neighbour searching strategies gave divergent
simulations due to the differing order in which forces were summed.

A further example is the handling of pseudo-random number generation
in Avida~\cite{ofria+wilke:2004}, an open source scientific software
platform for conducting and analysing experiments with
self-replicating and evolving computer programs. While it may
initially appear attractive to develop bespoke random number
generators within a system for consistency or performance across
platforms, this invariably adds complexity to your system and may
inhibit sharing and reproducibility.\\

\noindent {\textbf{Recommendation {\textrm{IX}}:}} Despite these
challenges to testing, unshared code is ultimately untestable.
Testing new complex scientific software is difficult -- until the
software is complete, unit tests may not be available. You should aim
to re-use modules or repos (git submodules) from publicly-shared
code; a corollary of Linus's Law (``given enough eyeballs, all bugs
are shallow'') might be that shared code is inherently more test-able.

% Recommendation X
\subsection{{\emph{Welcome to Web 2.0}}}

Virtual machines (VMs) in the cloud also make the testing of scaling
properties more simple.  If you have a tool that you claim is more
efficient, you could put together a cluster of slow nodes in the cloud
to demonstrate how well the software scales for parallel calculations.
Cloud computing is cheap, and getting cheaper. Algorithms that used to
require massive HPC resources can now be run cheaply by bidding on the
VM spot market. The web is a great leveller: use and share workflows
and web services~\cite{crick-et-al:2009b,oabarriaga-et-al:2014}.\\

\noindent {\textbf{Recommendation {\textrm{X}}:}} The web and the
cloud really do open up a whole new way of working. Even small,
seemingly trivial features like putting up a web interface to your
tool and its tests will allow users who are not able to install
necessary dependencies to explore the running of the tool
\cite{Hall2014}. Ultimately, this can lead to making an ``executable
paper'' appear on the Internet. The interactive {\em Try
F\#}(\url{http://www.tryfsharp.org/Learn}) and Z3
tutorials (\url{http://rise4fun.com/Z3/tutorial/guide}) are a
great start that begin to expose what can be done in this area.
% removed: ~\cite{tryFsharp} ~\cite{Z3tutorial}

\section{A Model for Reproducible Research Software}\label{sec:conclusion} 
%\section{Conclusions: A New Model}

Some of our Recommendations, such as ``{\emph{Be A Better Person}}''
or ``{\emph{The Lingua Franca}}'', are abstract, airy-fairy,
pie-in-the-sky even. However, most of them can be concretely realised
by a service for reproducibility. This service provides a concrete
implementation of free source code (``{\emph{Set The Code Free}}'')
that depends on other free source code (``{\emph{Lineage}}'') building
(``{\emph{YMMV}}'', ``{\emph{Welcome to Web 2.0}}'') and running tests
contributed in public (``{\emph{Data Representations}}'',
``{\emph{World Records}}'') in a completely reproducible regime.

The service we describe here can be seen as a specification. We have
not built it, but many services like travis-ci or Azure VSTS provide
some of the mechanical parts of it.  A service for reproducibility is
intended to play three important roles; it should:

\begin{enumerate}[i)]
\item Demonstrate that a piece of code can be compiled, run
and behaves as described, without manual intervention from the
developer;
\item Store and link specific artefacts with their linked
publications or other publicly-accessible datasets;
\item Allow new benchmarks to be added, by users other than
the developer, to widen the testing and identify potential bugs.
\end{enumerate}

The whole premise of our previous paper~\cite{crick-et-al_recomp2014}
is that {\emph{algorithms}} (and their {\emph{implementations}}) and
{\emph{models}} (sometimes also called {\emph{benchmarks}}) are
inextricably linked. Algorithms are designed for certain types of
models; models, though created to mimic some physical reality, also
serve to express the current known algorithms. An integrated autonomous
open cloud-based service can make this link explicit.

By developing a cloud-based, centralised service, which performs
automated code compilation, testing and benchmarking (with associated
auditing), we will link together published implementations of
algorithms and input models. This will allow the prototype to link
together software and data repositories, toolchains, workflows and
outputs, providing a seamless automated infrastructure for the
verification and validation of scientific models and in particular,
performance benchmarks. The program of work will lead the cultural
shift in both the short and long-term to move to a world in which
computational reproducibility helps researchers achieve their goals,
rather than being perceived as an overhead.

A system as described here has several up-front benefits: it links
research papers more closely to their outputs, making external
validation easier and allows interested users to explore unaddressed
sets of models. Critically, it helps researchers across computational
science to be more productive, rather than reproducibility being an
overhead on top of their day-to-day work. In the same way that tools
such as GitHub make collaborating easier while simultaneously allowing
effortless sharing, we envisage our system being similarly usable for
sharing and testing algorithms and their implementations, software,
models and benchmarks online.

Suppose you have come up with a better algorithm to deal with some
standard problem.  You write up the paper on the algorithm, and you
also push an implementation of your algorithm to the our cloud
environment's section on this standard problem. The effect of pushing
your implementation is to register your program as a possible
competitor in this standard problem competition. There exist several
dozen widely-agreed tests on this problem already on our cloud
environment's database. Maybe, after some negotiation due to your
novel approach to this standard problem, you add some of your own
tests to the database too.

\begin{figure}[!ht]
\centering
\includegraphics[width=0.9\columnwidth]{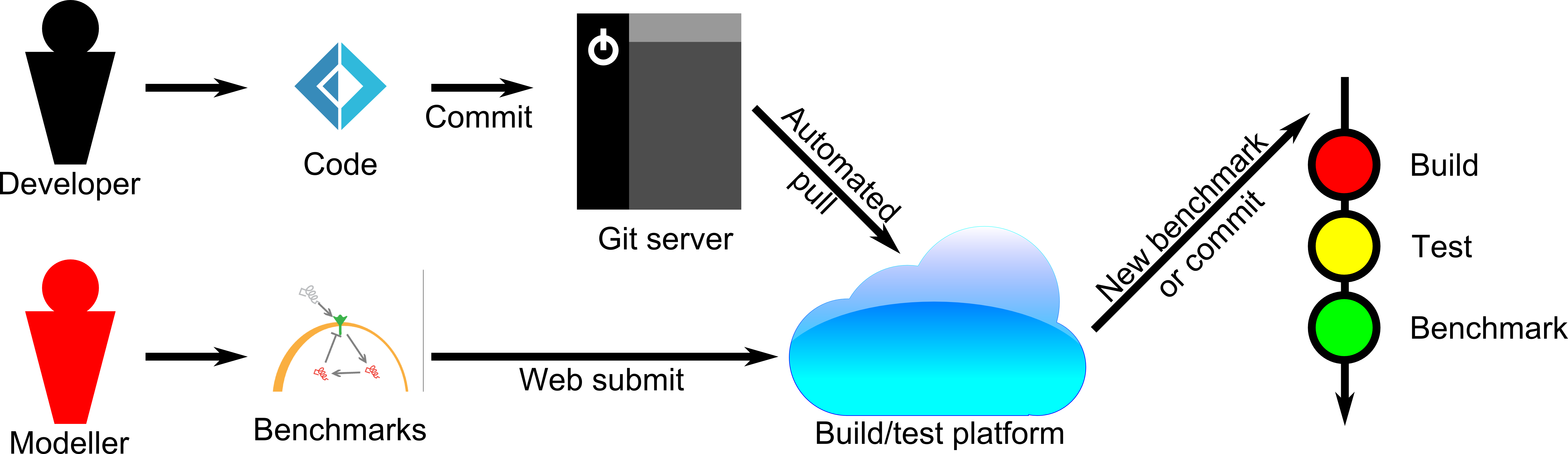}
\caption{Proposed reproducibility service workflow}
\label{fig:workflow} 
\end{figure}

Pushing your code activates the environment's continuous integration
system.  The cloud pulls in all the dependencies your code needs, on
the platforms you specify, and runs all the benchmarks. This happens
every time you push. It also happens every time one of your
dependencies (a library, a firmware upgrade for your platform, a new
API) changes too. This system (presented in Figure~\ref{fig:workflow})
would integrate with publicly available source code repositories,
automates the build, testing and benchmarking of algorithms and
benchmarks. It would allow testing models against competing
algorithms, and the addition of new models to the test suite (either
manually or from existing online repositories).

If we are truly serious about addressing the systemic socio-technical
issues in scientific disciplines that are underpinned by leveraging
software and computational techniques, then the proposal above would
bring together almost all of the points we have discussed in this
paper to provide an open research infrastructure for all. There are
already several web services that already aim to do many of these
things~\cite{rollins-et-al:2014,stodden-et-al:2015}, so a service that
can integrate most if not all of these features is possible. Such a
service would then allow algorithms and models to evolve together, and
be reproducible from the outset. Something more open and complete, and
stamped with the authority of the major domain
conferences/journals/national academies, would mean that your code
would never `bit-rot', and no one would have problems reproducing the
implementation of your published algorithm.

\section{Next Steps}

Following the proposal of such a system, the question becomes:
{\emph{how do we encourage widespread uptake, or even
standardisation?}}  Such a service would appear to be non-trivial,
given the large numbers of tools and workflows that could potentially
require to be supported by the service. After such a service has been
implemented, how do we ensure it is \emph{useful} and \emph{usable}
for researchers. Furthermore, how do we make it sustainable?

The benefits to the wider computational research community from a
cultural change to favour reproducibility are clear and as such we
should aim through software e-infrastructure and sharable,
community curated research workflows to mitigate these
costs. Furthermore, we can reasonably expect the distinct needs of
specific research communities to evolve over time, and initial
implementations of the platform may require refinement in response to
user feedback (supporting the critical cultural change by improving
the efficiency of researchers). As such, if the wider research
community is to move to requiring reproducibility, it seems most
reasonable that this is staggered over a number of years to allow for
both of these elements to develop, until eventually all researchers
are required to use the service.

The key question for different research communities then becomes:
{\emph{how to initialise this change?}} Such a requirement creates a
set of new costs to researchers, both in terms of time spent ensuring
that their tools work on the centralised system (in addition to their
local implementation), but also potentially in terms of equipment (in
terms of running the system). Such costs may be easier to bear for
some groups compared to others, especially those with large research
groups who can more easily distribute the tasks, and it is important
that the service does not present a barrier to early career
researchers and those with efficient budgets (this type of cost
analysis is not unique to reproducibility efforts -- it has been
estimated that a shift to becoming exclusively open access for a
journal may lead to a ten-fold increase in computer science
publication costs~\cite{vardi-cacm-2014}).

Nevertheless, this proposed new e-infrastructure could have a profound
impact on the way that computational science is performed,
repositioning the role of models, algorithms and benchmarks and
accelerating the research cycle, perhaps truly enabling a ``fourth
paradigm'' of data intensive scientific discovery~\cite{hey:2009}.
Ultimately though, continuing with an honest and open discussion of
what reproducibility means for the wider science research community is
important: we all need to explicitly confirm that this is worthwhile
and commit to addressing it, or don't bother doing it at all.

\section*{A Note on Re-Writing the WSSSPE Paper}

Many of the ideas, comments --- even attitudes --- in this paper come
from the authors' experience in programming, programming languages,
software. We have started from the Marc Andreessen comment that opens
this paper. In editing this paper from its original WSSSPE workshop
form, we realised that one assumption that seems to run through the
manuscript is that the behaviours we think are good are in fact those
that can be enforced in software.  Take mutability of variables in
programming as an example. Mutability increases the scope for bugs, so
modern programming languages like OCaml or C++14 enforce immutability
at the language or library level. But in fact immutability leads very
naturally to state-less or de novo build environments, and so to the
guideline that ``software must be compilable with de novo continuous
integration''.  And, similarly, so does the issue of openly publishing
your toolchain: it too must be compilable in a from-scratch build
environment to be of use to anyone else.

% BibTeX
\bibliographystyle{ieeetr}
\bibliography{jors2015}

\end{document}